\documentclass[aps,prb,twocolumn,letterpaper,showpacs,10pt]{revtex4-1}
\usepackage[utf8]{inputenc}
\usepackage{graphicx,amsmath}
\usepackage{amssymb}
\usepackage{amsthm}
\usepackage{hyperref}
\hypersetup{colorlinks=true, linkcolor=blue, citecolor=blue}

\newcommand{ \op }				{ {p} }
\newcommand{ \ox }				{ {x} }

\newcommand{ \ac }				[1] { \left\{{#1}\right\} }
\newcommand{ \com }			[1] { \left[ {#1} \right] }
\newcommand{ \ex }				[1] { \langle{#1}\rangle }

\newcommand{\pd}				{\phantom{\dag}}

\newcommand{\trm}				{\textrm}

\newcommand{\nn}				{\nonumber}

\newcommand{\ve}				{\varepsilon}
\newcommand{\fc}				{\Gamma}
\newcommand{\fd}				{\mathcal{D}}
\newcommand{\fg}				{\mathcal{G}}

\newcommand{\fl}				{\mathcal{L}}

\newcommand{ \tr }				{{\rm{Tr}}}

\begin{document}

\title{Teleportation-induced entanglement of two nanomechanical oscillators coupled to a topological superconductor}

\author{Stefan Walter$^1$}
\author{Jan Carl Budich$^{2,3,4}$}

\affiliation{
$^1$Department of Physics, University of Basel, Klingelbergstrasse 82, CH-4056 Basel, Switzerland \\
$^2$Department of Physics, Stockholm University, Se-106 91 Stockholm, Sweden\\
$^3$Institute for Theoretical Physics, University of Innsbruck, 6020 Innsbruck, Austria\\
$^4$Institute for Quantum Optics and Quantum Information,
Austrian Academy of Sciences, 6020 Innsbruck, Austria
}

\date{\today}

\pacs{03.65.Ud, 85.85.+j, 74.45.+c, 71.10.Pm}

\begin{abstract}
A one-dimensional topological superconductor features a single fermionic zero mode that is delocalized
over two Majorana bound states located at the ends of the system. We study a pair of spatially separated
nanomechanical oscillators tunnel-coupled to these Majorana modes. Most interestingly, we demonstrate
that the combination of electron-phonon coupling and a finite charging energy on the mesoscopic topological
superconductor can lead to an effective superexchange between the oscillators via the non-local fermionic
zero mode. We further show that this electron teleportation mechanism leads to entanglement of the two oscillators
over distances that can significantly exceed the coherence length of the superconductor.
\end{abstract}

\maketitle

\section{Introduction}
\label{sec:sec1}

In a 2001 paper,~\cite{Kitaev2001} Kitaev discovered the one-dimensional topological superconductor
(1DTSC) -- a one-dimensional proximity induced $p$-wave superconductor hosting a single Majorana
quasi particle (MQP) at each of its ends. More recently, experimentally feasible realizations of the 1DTSC
phase have been proposed \cite{LutchynTSC,OppenTSC} in semiconducting nanowires in proximity to an
$s$-wave superconductor. In these settings, the interplay of Rashba spin-orbit coupling and a magnetic
field induced Zeeman splitting in the nanowire gives rise to an effective $p$-wave pairing. By now, several
groups have reported first experimental signatures of MQPs in InSb nanowires.~\cite{LeoMaj,LarssonXu,HeiblumMaj}
Besides the fundamental interest attached to the experimental discovery of Majorana fermions in nature,
MQPs as realized in 1DTSC also have intriguing features relating to various aspects of fundamental
quantum physics: On the one hand the non-Abelian anyonic nature of MQPs shows great promise for
topological quantum information processing architectures.~\cite{Kitaev2001,Nayak:2008p51,Alicea:2011p260}
On the other hand the delocalized pair of MQPs at the ends of a 1DTSC can be viewed as a single
ordinary (spinless Dirac) fermionic zero mode leading to electron teleportation mechanisms,~\cite{Semenoff:2007p1479,Fu:2010}
i.e., coherent long-range quantum effects. In a hybrid system of a 1DTSC and two single-level quantum
dots, ground state entanglement of the occupation number of the quantum dots has been reported
in Ref.~\onlinecite{XuDots}.     

The understanding of genuine quantum effects on macroscopic lengthscales is one of the main motivations to
study nano-electromechanical systems~\cite{Poot} and nano-optomechanical systems.~\cite{AKM} In recent
years, decisive progress towards cooling nanomechanical resonators to the ground state has been reported.~
\cite{GroundStateCooling1,GroundStateCooling2,GroundStateCooling3,GroundStateCooling4}
However, long distance entanglement of nanomechanical systems which would be another experimental hallmark
in fundamental quantum physics has not been achieved yet although a variety of theoretical proposals have been
made.~\cite{Eisert:2004aa,Vitali:2007is,Cavities,MirrorMirrorEntanglement1,MirrorMirrorEntanglement2,MirrorMirrorEntanglement3,MirrorMirrorEntanglement4,Hammerer:2009tf,BEC1,BEC2,StefanJanJensBjoern}
In the interest of quantum coherence, different interaction mechanisms between spatially separated systems have
been suggested ranging from coupling to a common optical mode \cite{Cavities,MirrorMirrorEntanglement1,MirrorMirrorEntanglement2,MirrorMirrorEntanglement3,MirrorMirrorEntanglement4}
to exploiting the large coherence length of a Bose Einstein condensate \cite{BEC1,BEC2} and a Cooper-pair
condensate,~\cite{StefanJanJensBjoern} respectively. A hybrid system of a 1DTSC and one NEMO was
studied in Ref.~\onlinecite{WalterNEMSmsb}.

The article is organized as follows. In Sec.~\ref{sec:sec2}, we summarize our main results. We propose the setup, discuss
a possible realization of it, and introduce the Hamiltonian of the underlying model in Sec.~\ref{sec:sec3}. We present and
discuss the results of the generated entanglement in Sec.~\ref{sec:sec4}. Finally, we summarize in Sec.~\ref{sec:sec5}.

\begin{figure}[ht]
	\centering
	\includegraphics[width=0.95\columnwidth]{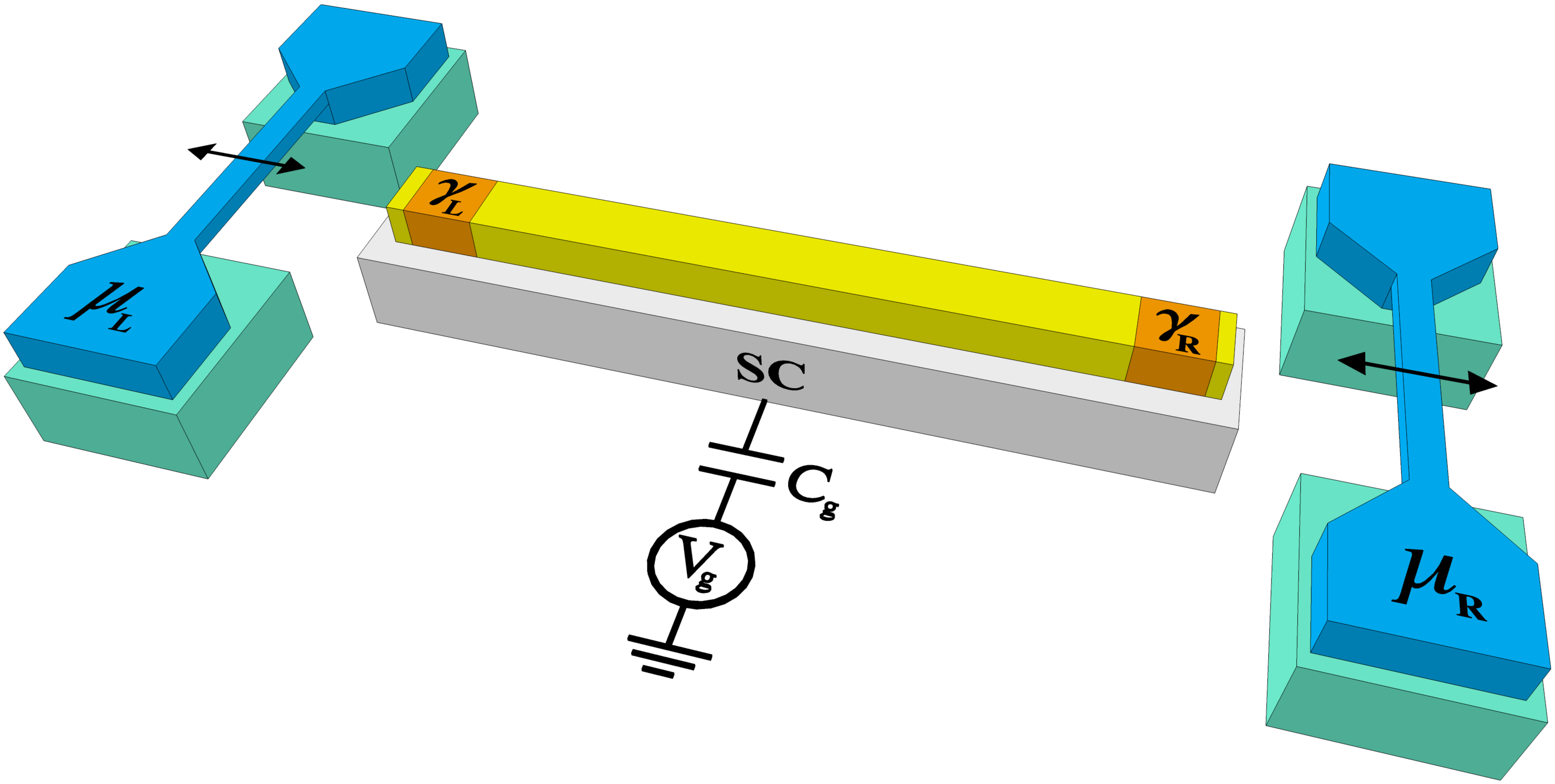}
	\caption{\label{fig:setup} (Color online) Schematics of the proposed setup. Two nano-electromechanical
	oscillators (blue) each tunnel coupled to one end of a one-dimensional topological superconductor. The
	one-dimensional topological superconductor is sketched as a nanowire (yellow) placed on top of a mesoscopic
	superconductor (gray). At each end of the wire a single Majorana quasi particle (orange) is located. A
	gate voltage $V_{g}$ is applied (across a gate capacitance $C_{g}$) to the mesoscopic superconductor 
	resulting in a finite charging energy $E_{c} = e^{2}/2 C_{g}$ of the superconductor. The nano-electromechanical
	oscillators are modeled as normal metal leads at the chemical potentials $\mu_{L/R}$.
	}
\end{figure}

\section{Main results}
\label{sec:sec2}

In this work, we bridge the research fields of topological superconductivity and entanglement in nanomechanical
systems by proposing a mechanism to entangle two nano-electromechanical oscillators (NEMOs).
More concretely, we demonstrate that the electron teleportation mechanism reported in Ref.~\onlinecite{Fu:2010} can
lead to an effective superexchange coupling of two distant NEMOs located in the vicinity of the opposite ends of
a mesoscopic 1DTSC. The combination of electron-phonon coupling on the NEMOs and a finite Coulomb charging
energy $E_{c} = e^{2}/2 C_{g}$ on the 1DTSC are shown to be the crucial ingredients for achieving long range
entanglement in the proposed setup.
The teleportation mechanism guarantees coherence at length scales that significantly exceed those of the
superconducting condensate wave function. In the proposed setup (see Fig. \ref{fig:setup}) entanglement between
two distant conducting NEMOs can be generated by simply driving a current through the device. Using a non-Markovian
master equation approach, we demonstrate that for NEMOs in their ground states, switching on a tunneling
current induces entanglement that persists over many oscillator periods. In the Markovian limit, we derive a Lindblad
master equation which provides an intuitive understanding of how number states of the NEMOs are dynamically
entangled by the superexchange coupling via the 1DTSC.

\section{Model}
\label{sec:sec3}

We will now show how an effective coupling between two NEMOs can be generated via an electron teleportation
mechanism involving the MQPs located at the ends of a 1DTSC. The proposed setup is shown in Fig.~\ref{fig:setup}
and is modeled by the following Hamiltonian (we put $\hbar$ $=$ $e$ $=$ $k_{B}$ $=$ $1$)
\begin{align}
	H &= \sum_{\alpha=L,R} H_{\trm{osc}}^{(\alpha)} + H_{\trm{lead}}^{(\alpha)} + H_{\rm{tun}} + H_{c}  \, , \nn
\end{align}
where
$
H_{\trm{osc}}^{(\alpha)} = \op_{\alpha}^{2}/2 m_{\alpha} +m_{\alpha} \Omega_{\alpha}^{2} \ox_{\alpha}^{2}/2
$
describes the two NEMOs denoted by $\alpha=L,R$~ with effective mass $m_{\alpha}$, frequency $\Omega_{\alpha}$,
and position and momentum operators $\ox_{\alpha}$ and $\op_{\alpha}$, respectively. For simplicity, we assume that
$m_{L}=m_{R}$~and $\Omega_{L}=\Omega_{R}$. The conducting NEMOs act as two independent normal metal leads
which are characterized by the Hamiltonians
$
H_{\trm{lead}}^{(\alpha)} = \sum_{k} \ve_{k}^{\pd} \psi_{\alpha k}^{\dag} \psi_{\alpha k}^{\pd}
$
and which are held at the chemical potentials  $\mu_{L/R}$. The tunneling Hamiltonian $ H_{\rm{tun}}$~ from a normal
metal lead into a 1DTSC without charging energy can be written as~\cite{Bolech:2007}
\begin{align}\label{eqn:mo2}
	H_{\rm{tun}} 	= \sum_{k} [ i T^{\pd}_{L} ( \psi^{\pd}_{Lk} + \psi^{\dag}_{Lk} ) \gamma_{L}^{\pd} + (L \to R) ] \, ,
\end{align}
where, in general, the tunneling amplitudes $T_{\alpha}$ have an exponential dependence on the displacement
of the NEMOs, i.e., $T_{\alpha} \sim e^{-x_{\alpha}/x_{0}}$. As the oscillation amplitude is assumed to be small compared to the
mean distance between the edge of the 1DTSC and the NEMO, we approximate $T_{\alpha}$ to depend linearly
on the oscillator displacement: $T_{\alpha} = t_{0\alpha} + t_{x\alpha} \ox_{\alpha}$. Such a tunneling gap between a
suspended gold beam and an electronic reservoir was realized in Ref.~\onlinecite{Flowers:2007}. Other possibilities
include, for instance, replacing the suspended metallic beam by a vibrating metallic tip or by a shuttle-like device.~\cite{shuttle}
As yet another possibility to achieve such a coupling, the suspended point contacts could be replaced by
an electrostatically gated connection to the 1DTSC that is modulated piezoelectrically or capacitively by
the NEMO.

\begin{figure}[t]
 	\centering
 	\includegraphics[width=0.95\columnwidth]{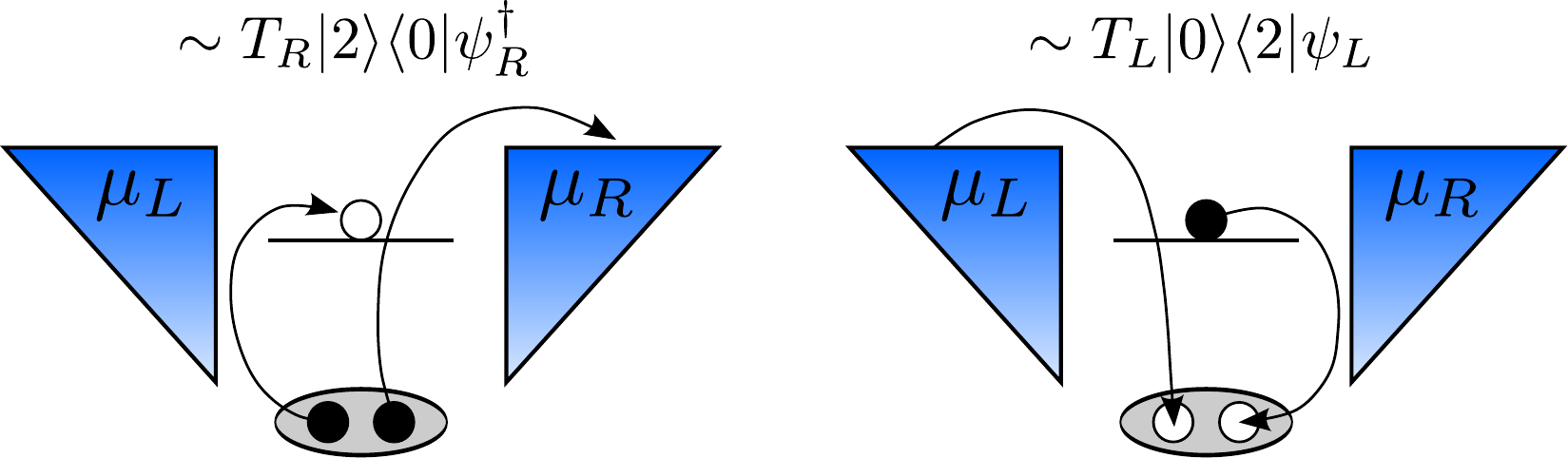}
 	\caption{\label{fig:anomalouscot} (Color online) Anomalous cotunneling process decomposed into two
	(virtual) steps. Looking only at input and output states, one electron tunnels from the left to the right lead (blue). The
 	state of the superconductor is unchanged. This is the only anomalous second order process contributing
	to the current that does not contain states with $E>E_{c}$. The oval denotes the a Cooper pair in the
 	condensate. The level in the middle is the subgap-fermion $c$. The tunnel processes depicted here
	appear in the effective tunnel Hamiltonian Eq.~(\ref{eqn:mo3}).
 	}
 \end{figure}

The left ($\gamma_{L}$) and right ($\gamma_{R}$) MQP satisfy $\left\{ \gamma_{i}, \gamma_{j} \right\} = 2 \delta_{ij}$
and can be expressed as
$
\gamma_{L} = (c^{\pd} + c^{\dag})
$
and
$
\gamma_{R} = -i(c^{\pd} - c^{\dag})
$,
where $c$ and $c^{\dag}$ are the annihilation and creation operators, respectively, of a single spinless Dirac fermion that is
delocalized over the two ends of the 1DTSC. Equation~(\ref{eqn:mo2}) contains so called anomalous terms which
break particle number conservation in the mean field picture of superconductivity as they microscopically involve
the creation or annihilation of a Cooper pair which is not explicitly accounted for at that level of description.
In a 1DTSC with zero charging energy $E_{c}=0$, the NEMOs independently couple locally to the two
ends of the 1DTSC and the effective coupling necessary for entangling the oscillators is absent. However, the
situation is different in a mesoscopic superconductor with a finite charging energy $E_{c}$~which gives rise to
an explicit dependence of the energy on the number of electrons. Hence, one has to go beyond the effective
description of Eq. (\ref{eqn:mo2}) and explicitly keep track of the change in the number of Cooper pairs in the
condensate during anomalous tunneling processes. The gate voltage $V_{g}$~is assumed to be adjusted such
that the number of Cooper pairs $N_C$~ in the ground state of the 1DTSC is $N_0$~and the occupation number
$n_c=c^\dag c$~of the delocalized fermionic bound state is zero. The charging Hamiltonian $H_{c}$ then reads
$
H_{c} = E_{c} \left(2 N_{C} + n_{c} - 2 N_{0} \right)^{2}
$.
We would like to point out that $n_{c} = \frac{1}{2}( i \gamma_{L} \gamma_{R} + 1)$~as appearing in $H_c$
effectively couples the two MQPs $\gamma_L$~and $\gamma_R$~even if the direct overlap of the two bound
state wave functions is negligible. This coupling is crucial for the electron teleportation mechanisms as it prevents the
dynamical independence of the two MQPs. We would like to focus on the parameter regime $T_{\alpha}, V < \Omega_{\alpha} < E_{c} < \Delta \to \infty$.
In this limit, non-local tunneling processes involving continuum states of the superconductor (e.g. Crossed
Andreev Reflection or electron cotunneling) are suppressed. Moreover, in this scenario, there are no
resonant levels in the superconductor for first-order tunneling processes. However,
second-order cotunneling processes via virtual states with energies on the order of $E_c$~are allowed and lead
to an effective superexchange coupling between the NEMOs as we will derive now. We neglect processes
containing intermediate states with two or more excess electrons on the superconductor which are suppressed
by an energy denominator of at least $4 E_{c}$~and are hence less relevant. This approximation excludes
all terms where an extra Cooper pair is created. The only anomalous second order tunneling process which
is then allowed is the anomalous cotunneling depicted in Fig. \ref{fig:anomalouscot}.

After truncating the Hilbert space of the superconductor to the eigenstates with $E\le E_c$, we obtain a three-dimensional
Hilbert space with the basis
\begin{align*}
	&\lvert 0\rangle=\lvert N_C=N_0,n_c=0\rangle		&&E_{0}=0 \\
	&\lvert 1\rangle=\lvert N_C=N_0,n_c=1\rangle		&&E_{1}=E_{c}\\
	&\lvert 2\rangle=\lvert N_C=N_0-1,n_c=1\rangle	&&E_{2}=E_{c}
\end{align*}
In this basis, $H_c$~can be represented as:
$
H_{c} = \trm{diag}\{ 0, E_{c}, E_{c} \}
$.
The tunneling Hamiltonian (\ref{eqn:mo2}) constrained to the truncated Hilbert space of the superconductor reads as
\begin{align}\label{eqn:mo3}
	H_{\rm{tun}} \approx~ & iT_{L} \sum_{k} \left(\lvert 1 \rangle \langle 0 \rvert \psi_{Lk} - \lvert 2 \rangle \langle 0 \rvert \psi_{Lk}^{\dag} \right) \nn \\
 					  +~&T_{R} \sum_{k} \left(-\lvert 1 \rangle \langle 0 \rvert \psi_{Rk} + \lvert 2 \rangle \langle 0 \rvert \psi_{Rk}^{\dag} \right) + \text{h.c.}.
\end{align}
The terms in Eq. (\ref{eqn:mo3}) which involve the breaking and recombination of a Cooper pair, respectively are illustrated in Fig. \ref{fig:anomalouscot}.
Assuming that the superconductor is initially in its ground state $\lvert 0\rangle$, we can integrate out the first order
tunnel coupling to the excited states $\lvert 1\rangle,\lvert 2\rangle$. That way, we obtain an effective direct tunneling
Hamiltonian between the left and the right lead containing the leading second order cotunneling processes in the
original tunnel coupling Eq.~(\ref{eqn:mo2}). Explicitly, we get
\begin{align}\label{eqn:mo4}
	H_{\trm{tun}}^{(\trm{eff})} = -\frac{T^{2}_{L} + T^{2}_{R}}{E_{c}} - \frac{2T_{L} T_{R}}{E_{c}} \sum_{k} [ i \psi_{Lk}^{\dag} \psi^{\pd}_{Rk} + \trm{h.c.} ] \, .
\end{align}
Recalling the position dependence $T_{\alpha} = t_{0\alpha} + t_{x\alpha} \ox_{\alpha}$~of the tunnel couplings, it
becomes clear that Eq. (\ref{eqn:mo4}) also contains an effective direct coupling between the NEMOs.
This formally mimics the superexchange coupling which could also be achieved using a single quantum dot with a finite charging
energy. However, we would like to stress two conceptual advantages of the electron teleportation-induced superexchange coupling.
First, it guarantees phase coherent coupling between the NEMOs over distances where the confinement induced level spacing on a
quantum dot would become very small. Second, the tunneling density of states associated with the delocalized fermion $c$~in our
setting is spatially strongly peaked around the interface between the NEMO and the 1DTSC. In a large single level quantum dot in
contrast, the same spectral weight would be smeared out all over the "bulk" of the dot. In the following, we will demonstrate how
this teleportation-induced superexchange coupling can be employed to generate entanglement between the oscillators over
distances which are not limited by the coherence length of the superconducting condensate.

\section{Entanglement}
\label{sec:sec4}

As shown above (see Eq. (\ref{eqn:mo4})), tunnel coupling two NEMOs to a 1DTSC leads to an effective direct
coupling between the NEMOs. Therefore, we expect the generation of entanglement in the bipartite continuous variable
system consisting of the two NEMOs. We study the time evolution of entanglement between the two NEMOs using the logarithmic negativity as
an entanglement measure: $E_{N}(\rho_{\trm{osc}}) = \log_{2}(\| \rho_{\trm{osc}}^\Gamma\|_{1})$.~
\cite{Negativity1,Negativity2,Negativity3} Here, $\rho_{\trm{osc}}^\Gamma$ is the partial transpose of the state
of the bipartite system. For a Gaussian state, the logarithmic negativity can be computed from the covariance matrix
$\fc_{j,k}(t) = \tr[ \rho_{\trm{osc}} (t) \{R_j, R_k \} ]$, where ${R} = (\ox_{1},\op_{1},\ox_{2},\op_{2})^{T}$ is the vector of
quadratures. We compute the time dependence of the entries of $\fc(t)$ by solving the equation of motion for the system's
density matrix $\rho_{\trm{osc}}(t)$ employing a time convolutionless master equation method.~\cite{Breuer:2002wp}
Within our effective tunneling Hamiltonian approach (see Eq. (\ref{eqn:mo4})), the master equation in the Born approximation
is given by
\begin{align}\label{eqn:nemsEnt4}
	\dot{\rho}_{\trm{osc}}(t) =	&-i \com{H_{\trm{osc}},\rho_{\trm{osc}}(t)}  \\
					- \int_{0}^{t} d\tau \, & \tr_{\trm{leads}} \left[ H_{\trm{tun}}^{(\trm{eff})}, \left[ H_{\trm{tun}}^{(\trm{eff})}(\tau-t),\rho_{\trm{osc}}(t) \otimes \rho_{\trm{leads}} \right]  \right] \, . \nn
\end{align}
For the sake of simplicity, we assume in the following identical NEMOs ($\Omega_{\alpha}=\Omega$ and
$m_{\alpha}=m$). We also chose a symmetric coupling and real tunneling amplitudes ($t_{0\alpha} = t_{0}$
and $t_{x\alpha} = t_{x}$).

\begin{figure}[t]
	\centering
	\includegraphics[width=0.85\columnwidth]{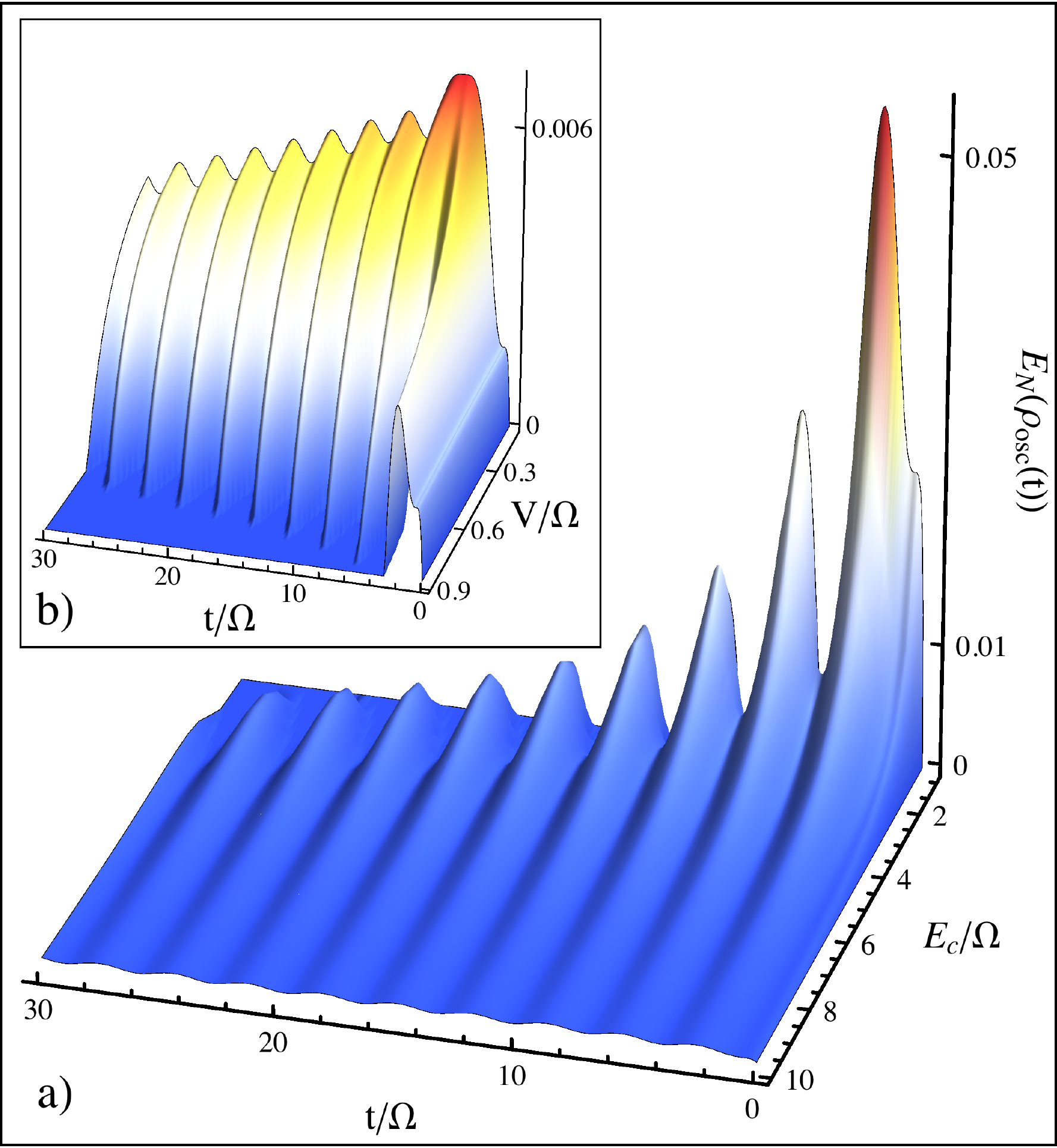}
	\caption{\label{fig:Enall} (Color online) a) Logarithmic negativity as a function of $E_{c}$ and time for
	$V/\Omega=0.5$. For small $E_{c}$ the generated entanglement is higher but decays faster than for large
	$E_{c}$. b) Logarithmic negativity as a function of $V$ and time for $E_{c}/\Omega = 4$. For lower bias
	voltages the entanglement is higher. In both cases, the other parameters are $T_{el} = 0$, $L_{c}/\Omega=1$,
	and $t_{0} t_{x}/\sqrt{m \Omega} = 0.1$.
	}
\end{figure}

Up to second order in $t_x$, i.e., only taking into account terms $\sim (t_{0} t_{x})^{2}$ in Eq.~(\ref{eqn:nemsEnt4}),
the time dependence of the covariance matrix $\fc(t)$ can be obtained similarly as in Ref.~\onlinecite{StefanJanJensBjoern},
for technical details we refer to the Appendix.
In Fig.~\ref{fig:Enall}, we show results for the logarithmic negativity, taking for simplicity the vacuum state
as an initial state. The Gaussian character of this initial state is preserved at all times of the dynamics.
Figure~\ref{fig:Enall}a) shows the time dependence of the logarithmic negativity $E_{N}$ for a fixed bias
voltage $V = \mu_{L} - \mu_{R}$ and for various values of the charging energy $E_{c}$. We see that the
two NEMOs become entangled right after the tunneling has been suddenly switched on. The generated
entanglement is higher but decays faster for smaller values of $E_{c}$ compared to larger values of $E_{c}$.
Figure~\ref{fig:Enall}b) shows $E_{N}$ over time for a fixed charging energy $E_{c}$ for different bias voltages $V$.
Here, we see that lower voltages lead to a higher logarithmic negativity. This can be interpreted by recognizing
that the bias voltage is similar to an effective temperature of the leads and thereby leads to decoherence.
As a first result, we conclude that an effective interaction mediated by an electron teleportation mechanism
involving MQPs leads to the generation of entanglement of two distant NEMOs.
To lowest order in tunneling $(t_{0} t_{x})^{2}$, the entanglement is due to damping and decoherence
mechanisms described by time-dependent kernels $\fg(t)$, cf. Appendix.
However, the effective tunneling Hamiltonian Eq.~(\ref{eqn:mo4}) together with the equation of motion
for $\rho_{\trm{osc}}$ leads to contributions of order $t_{x}^{4}$ in the equation of motion for the NEMOs.

Next, we analyse exactly these contributions. Restricting ourselves to the low-bias limit, we show that
entanglement between the NEMOs can be generated in a purely dissipative fashion, described by a Lindblad
master equation. In the limit of low-bias voltages, it is not possible to excite any of the NEMOs by the applied
bias voltage. This allows us to employ the rotating wave approximation, i.e., excitations can only
be interchanged between the two NEMOs. In the low-bias limit and taking the Markovian limit, the equation
of motion reduces to
\begin{align}
	&\dot \rho_{\trm{osc}} = \fl [\rho_{\trm{osc}}] = -i [H_{\trm{osc}}, \rho_{\trm{osc}}] + \gamma \mathcal{D}[O] \rho_{\trm{osc}} \nn 
\end{align}	
with the Liouvillian superoperator  $\fl$ and a Lindblad dissipator $\fd[O]\rho = O \rho O^{\dag} - \frac{1}{2} \left\{ O^{\dag} O, \rho \right\}$.
In our case we have $\gamma = \frac{\pi \, t_{x}^{4}}{E_{c}^{2}} \frac{\rho_{L} \rho_{R}}{(m \Omega)^{2}} V$~and
$O = {a}_{L}^{\dag} {a}_{R}^{\pd} + {a}_{R}^{\dag}{a}_{L}^{\pd}$, where
${a}_{\alpha} = (\ox_{\alpha} \sqrt{m \Omega} + i \op_{\alpha}/\sqrt{m\Omega} ) /\sqrt{2}$ and 
${a}_{\alpha}^{\dag} = (\ox_{\alpha} \sqrt{m \Omega} - i \op_{\alpha}/\sqrt{m\Omega} ) /\sqrt{2}$ are bosonic
annihilation and creation operators, respectively. $\rho_{\alpha}$ is the density of states in lead $\alpha$ which
we assume as constant in the relevant energy window.
If other dissipation channels such as an additional bosonic heat bath are absent, the steady state of
the system ($\fl [\rho_{ss}] = 0$) is not unique. However, if the number of excitations
($N_{\trm{tot}}$ = $\sum_{\alpha} {a}_{\alpha}^{\dag} {a}_{\alpha}^{\pd}$ = $\sum_{\alpha} n_{\alpha}$) is kept fixed,
the steady state is unique. For instance, the pure initial state $\lvert \Psi \rangle = \lvert n_{L}=1,n_{R}=1 \rangle$ is
dissipatively driven to the (mixed) entangled state
\begin{align}
	\rho_{ss} = \frac{1}{2} \lvert \Psi \rangle \langle \Psi \rvert + \frac{1}{2} \lvert \Phi \rangle \langle \Phi \rvert \, , \nn
\end{align}
where $\lvert \Phi \rangle = \frac{1}{\sqrt{2}} ( \lvert 2,0 \rangle + \lvert 0,2 \rangle)$ is a maximally entangled state.
The degree of entanglement of $\rho_{ss}$ is readily quantified by calculating $E_{N}(\rho_{ss}) = \log_{2}(3/2)$.
In the presence of a finite temperature heat bath, sectors of different particle number will start to couple. Thereby,
the stationary state becomes unique and entanglement is unsurprisingly lost. However, processes destroying and
generating the entanglement now compete with each other. This still allows for the generation of entanglement
in a dissipation fashion. The rates of the entanglement generating and destroying processes (characterized by an independent
rate determined by the microscopic environment of the NEMOs) are governed by their respective Liouvillian gaps, for
details we also refer to Ref.~\onlinecite{StefanJanJensBjoern}.

\section{Concluding discussion}
\label{sec:sec5}

To summarize, we have shown that entanglement between two distant NEMOs can be achieved by tunnel coupling
of the NEMOs to two MQPs residing at the ends of a 1DTSC. A finite charging energy on the 1DTSC leads to an
effective superexchange coupling between the NEMOs via the non-local MQPs. This electron teleportation mechanism
guarantees phase coherence over length scales $\sim 1/E_c$~ that are significantly larger than the superconducting
coherence length $\sim 1/\Delta$. Our proposal allows for entangling two mesoscopic NEMOs initially cooled to their
ground states in an all electronic setup by driving a current through the device. In the Markov approximation, the
equation of motion for the system's density matrix $\rho_{\trm{osc}}$ reduces to a Lindblad master equation. In this
limit, NEMOs initially prepared in number states can be entangled by purely dissipative means.

We briefly want to elaborate on the conceptual difference between our work and Ref.~\onlinecite{XuDots}, where the non-local
nature of a pair of MQPs was exploited to create a charge-entangled ground state of two single level quantum dots in
the Coulomb blockade regime. On the contrary, in our proposal, the electron charge degrees of freedom are in fact
only used to generate an effective superexchange coupling between two rather \emph{macroscopic mechanical
degrees of freedom}. In our setting, entanglement is not a ground state property of a closed system but is dynamically
generated by driving a current between the two metallic leads. Remarkably, the thermalization (decoherence) of the
electrons after their tunneling into these reservoirs does not affect the coherence times of the entangled NEMOs.

Our analysis relies crucially on the hierarchy $V,T_{\alpha}<\Omega<E_c<\Delta$~of the involved energy scales. Finally, we
would like to discuss experimentally relevant energy scales in the proposed setup thereby demonstrating the feasibility
of the assumed parameter regime. For an $\trm{InSb}$ wire proximity coupled to a $\trm{NbTiN}$ superconductor, experimental
data reported in Ref.~\onlinecite{LeoMaj} indicate an induced gap on the order of $\Delta = 250 \, \mu eV$. By varying the
size of the superconductor, the charging energy $E_{c}$ can be adjusted. Here, we assume $E_{c} = 20 \, \mu eV$. Frequencies
of doubly clamped NEMOs can be as high as $\Omega = 500 \, \trm{MHz}\approx 2\,\mu eV$,~\cite{Li2008} i.e., one
order of magnitude smaller than a typical charging energy. Still, such NEMOs could be passively cooled to their
ground state at typical dilution refrigerator temperatures. Taking these estimates, the localization length of the MQPs
at the ends of the 1DTSC is about $2\, \mu m$. For the assumed charging energies, the MQPs could be separated
by at least $20\, \mu m$, hence direct tunneling between them is negligible.

\begin{acknowledgements}
We would like to thank Christoph Bruder, Patrik Recher, Thomas Schmidt, and Bj{\"o}rn Trauzettel for stimulating
discussions. SW acknowledges financial support form the Swiss SNF and the NCCR Quantum Science and Technology.
JCB acknowledges financial support from the Swedish Research Council (VR) and the ERC Synergy Grant UQUAM.
\end{acknowledgements}

\appendix*
\section{Details on the equation of motion}

In this Appendix, we give details on the equation of motion of the two NEMOs. For simplicity, we assume identical NEMOs ($\Omega_{\alpha}=\Omega$,
$m_{\alpha}=m$) and symmetric coupling ($t_{0\alpha} = t_{0}$, $t_{x\alpha} = t_{x}$). Using the effective tunneling Hamiltonian, Eq.~(\ref{eqn:mo4}) of
the main text, the equation of motion, Eq.~(\ref{eqn:nemsEnt4}) of the main text, can be written as 
\begin{align}
	\dot{\rho}_{\trm{osc}}(t) = &-i\com{H_{\trm{osc}} + i \fg_{-}^{(c)}(t) ({x}_{L} + {x}_{R})^{2} , \rho_{\trm{osc}}(t)} \nn \\
				&- \fg_{+}^{c}(t) \com{{x}_{L} + {x}_{R}, \com{{x}_{L} + {x}_{R}, \rho_{\trm{osc}}(t)} } \nn \\
				&+ \fg_{+}^{s}(t) \com{{x}_{L} + {x}_{R}, \com{{p}_{L} + {p}_{R}, \rho_{\trm{osc}}(t)} } \nn \\
				&+ \fg_{-}^{s}(t) \com{{x}_{L} + {x}_{R}, \ac{{p}_{L} + {p}_{R}, \rho_{\trm{osc}}(t)} } \nn \\
				&- \fg_{+}^{cc}(t) \com{{x}_{L} {x}_{R}, \com{{x}_{L} {x}_{R}, \rho_{\trm{osc}}(t)} } \nn \\
				&- \fg_{+}^{ss}(t) \com{{x}_{L} {x}_{R}, \com{{p}_{L} {p}_{R}, \rho_{\trm{osc}}(t)} } \nn \\
				&+ \fg_{+}^{cs}(t) \com{{x}_{L} {x}_{R}, \com{{x}_{L} {p}_{R}+{x}_{R} {p}_{L}, \rho_{\trm{osc}}(t)} } \nn \\
				&- \fg_{-}^{cc}(t) \com{{x}_{L} {x}_{R}, \ac{{x}_{L} {x}_{R}, \rho_{\trm{osc}}(t)} } \nn \\
				&- \fg_{-}^{ss}(t) \com{{x}_{L} {x}_{R}, \ac{{p}_{L} {p}_{R}, \rho_{\trm{osc}}(t)} } \nn \\
				&+ \fg_{-}^{cs}(t) \com{{x}_{L} {x}_{R}, \ac{{x}_{L} {p}_{R}+{x}_{R} {p}_{L}, \rho_{\trm{osc}}(t)} }  \nn \, .
\end{align}

Non-Markovian effects are included in the equation of motion by the time-dependent kernels given by
\begin{align}
	\fg_{+}^{c}(t)		&= \int_{0}^{t} d\tau \left( G^{(1)}(\tau) + G^{(1)}(-\tau) \right) (2 t_{0} t_{x})^{2} \cos(\Omega \tau) \, , \nn \\
	\fg_{+}^{s}(t)	&= \int_{0}^{t} d\tau \left( G^{(1)}(\tau) + G^{(1)}(-\tau) \right) \frac{(2 t_{0} t_{x})^{2}}{m \Omega} \sin(\Omega \tau) \, , \nn \\
	\fg_{-}^{(c)}(t) 		&= \int_{0}^{t} d\tau \left( G^{(2)}(\tau) - G^{(2)}(-\tau) \right) (2 t_{0} t_{x})^{2} \cos(\Omega \tau) \, , \nn \\
	\fg_{-}^{s}(t)	&= \int_{0}^{t} d\tau \left( G^{(2)}(\tau) - G^{(2)}(-\tau) \right) \frac{(2 t_{0} t_{x})^{2}}{m \Omega} \sin(\Omega \tau) \, , \nn
\end{align}	
\begin{align}	
	\fg_{+}^{cc}(t) = \int_{0}^{t} d\tau &\left( G^{(1)}(\tau) + G^{(1)}(-\tau) \right) 4 t_{x}^{4} \cos(\Omega \tau) \cos(\Omega \tau) \, , \nn \\
	\fg_{+}^{ss}(t) = \int_{0}^{t} d\tau &\left( G^{(1)}(\tau) + G^{(1)}(-\tau) \right) \frac{4 t_{x}^{4}}{m^{2} \Omega^{2}} \sin(\Omega \tau) \sin(\Omega \tau) \, ,\nn \\
	\fg_{+}^{cs}(t) = \int_{0}^{t} d\tau &\left( G^{(1)}(\tau) + G^{(1)}(-\tau) \right) \frac{4 t_{x}^{4}}{m \Omega} \cos(\Omega \tau) \sin(\Omega \tau) \, , \nn \\
	\fg_{-}^{cc}(t) = \int_{0}^{t} d\tau &\left( G^{(2)}(\tau) - G^{(2)}(-\tau) \right) 4 t_{x}^{4} \cos(\Omega \tau) \cos(\Omega \tau) \, , \nn \\
	\fg_{-}^{ss}(t) = \int_{0}^{t} d\tau &\left( G^{(2)}(\tau) - G^{(2)}(-\tau) \right) \frac{4 t_{x}^{4}}{m^{2} \Omega^{2}} \sin(\Omega \tau) \sin(\Omega \tau) \, , \nn \\
	\fg_{-}^{cs}(t) = \int_{0}^{t} d\tau &\left( G^{(2)}(\tau) - G^{(2)}(-\tau) \right) \frac{4 t_{x}^{4}}{m \Omega} \cos(\Omega \tau) \sin(\Omega \tau) \, .\nn
\end{align}
The functions $G^{(1)}(t)$ and $G^{(2)}(t)$ are given by
\begin{align}
	G^{(1)}(t) &= \frac{1}{2} \ex{\ac{B(t),B^{\dag}(0)}}	\nn \, ,\\
	G^{(2)}(t) &= \frac{1}{2} \ex{\com{B(t),B^{\dag}(0)}} 	\nn \, ,
\end{align}
with
\begin{align}
	B = -\frac{i}{E_{c}} \psi^{\dag}_{L} \psi^{\pd}_{R} \, . \nn
\end{align}
With this we obtain
\begin{align}
	&G^{(m)}(t) = \frac{1}{2 E_{c}^{2}} \int d\ve_{L} \int d\ve_{R} \, J(\ve_{L},\ve_{R}) \, e^{i (\ve_{L}-\ve_{R})t}  \nn \\ 
		&\times \big[ n_{L}(\ve_{L}) (1-n_{R}(\ve_{R})) - (-1)^{m} n_{R}(\ve_{R}) (1-n_{L}(\ve_{L})) \big] \, ,\nn
\end{align}
where $n_{x}(\ve_{x}) = (e^{\beta(\ve_{x}-\mu_{x})}+1)^{-1}$ is the Fermi distribution function (with $\beta = 1/T_{el}$ being the
inverse electronic temperature of the leads; we set $k_{B} = 1$) and
\begin{align}
	J(\ve_{L},\ve_{R}) = \sum_{k,q} \delta(\ve_{L}-\ve_{k}) \delta(\ve_{R}-\ve_{q}) \, , \nn
\end{align}
is an energy-dependent spectral function. To account for a finite lifetime of quasiparticles in the leads, the $\delta$-functions
are smeared out and replaced by Lorentzians of width $L_{c}$
\begin{align}
	J(\ve_{L},\ve_{R}) =\sum_{k,q} \frac{L_{c}}{(\ve_{L}-\ve_{k})^{2}+L_{c}^{2}} \frac{L_{c}}{(\ve_{R}-\ve_{q})^{2}+L_{c}^{2}} \, . \nn
\end{align}

Energies close to the Fermi level of each lead will contribute most to each of the independent sums. To keep the number
of parameters as low as possible, we restrict ourselves to the regime of low applied bias voltages ($V<L_{c}$). Then, we
can be approximate the energy-dependent spectral function as
\begin{align}
	J(\ve_{L},\ve_{R}) = \frac{1}{(\ve_{L}-\ve_{R})^{2}+L_{c}^{2}} \, , \nn
\end{align}
which implies that an electron with energy $\ve_{l}$ in the left lead can tunnel into states of the right lead with energy $\ve_{r}$,
broadened by $L_{c}$~.\cite{WingreenEtAl1,WingreenEtAl2,WingreenEtAl3,Chen}
The limit $L_{c} \rightarrow 0$ resembles a resonant tunneling process with narrow densities of states in the leads. The
opposite limit, $L_{c} \rightarrow \infty$, corresponds to the so-called wide-band limit with an energy-independent density
of states in the leads, i.e., any electron from the left lead can tunnel into the right lead. With this, all the above kernels
can be calculated analytically. The resulting expressions are not very insightful and too lengthy to be stated here.

\bibliographystyle{apsrev}

\begin{thebibliography}{99}

\bibitem{Kitaev2001}
	A.\ Kitaev,
  	Physics-Uspekhi {\bf 44}, 131 (2001).

\bibitem{LutchynTSC}
	R.\ M.\ Lutchyn, J.\ D.\ Sau, and S.\ Das Sarma,
	Phys.\ Rev.\ Lett.\ {\bf 105}, 077001 (2010).
 
\bibitem{OppenTSC}
	Y.\ Oreg, G.\ Refael, and F.\ von Oppen,
	Phys.\ Rev.\ Lett.\ {\bf 105}, 177002 (2010).
	
\bibitem{LeoMaj}
	V.\ Mourik, K.\ Zuo, S.\ M.\ Frolov, S.\ R.\ Plissard, E.\ P.\ A.\ M.\ Bakkers, and L.\ P.\ Kouwenhoven,
	Science {\bf 336}, 1003 (2012).
 
\bibitem{LarssonXu}
 	M.\ T.\ Deng, C.\ L.\ Yu, G.\ Y.\ Huang, M.\ Larsson, P.\ Caroff, and H.\ Q.\ Xu,
	Nano Lett.\ {\bf 12}, 6414 (2012).
 
\bibitem{HeiblumMaj}
 	A.\ Das, Y.\ Ronen, Y.\ Most, Y.\ Oreg, M.\ Heiblum, and H.\ Shtrikman,
	Nat. Phys. {\bf 8}, 88 (2012).

\bibitem{Nayak:2008p51}
	C.\ Nayak, A.\ Stern, M.\ Freedman, and S.\ Das Sarma. 
	Rev.\ Mod.\ Phys. {\bf 80}, 1083 (2008).

\bibitem{Alicea:2011p260}
	J.\  Alicea, Y.\ Oreg, G.\ Refael, F.\ von Oppen, and M.\ P.\ A.\ Fisher,
	Nat. Phys. {\bf 7}, 412 (2011).
   
\bibitem{Semenoff:2007p1479}
	G.\ Semenoff and P.\ Sodano,
	J. Phys. B {\bf 40}, 1479 (2007).
 
\bibitem{Fu:2010}
 	L.\ Fu,
	Phys.\ Rev.\ Lett.\ {\bf 104}, 056402 (2010).
	
\bibitem{XuDots}	
	Z.\ Wang, X.-Y.\ Hu, Q.-F.\ Liang, and X.\ Hu,
	Phys.\ Rev.\ B {\bf 87}, 214513 (2013).	
 
\bibitem{Poot}
	M.\ Poot and H.\ S.\ J.\ van der Zant,
	Phys.\ Rep.\ {\bf 511}, 273 (2012).		

\bibitem{AKM}
	M.\ Aspelmeyer, T.\ J.\ Kippenberg, and F.\ Marquardt,
	 arXiv:1303.0733 (2013)

\bibitem{GroundStateCooling1}
	A.\ D.\ O'Connell, M.\ Hofheinz, M.\ Ansmann, R.\ C.\ Bialczak, 	M.\ Lenander, E.\ Lucero, M.\ Neeley, D.\ Sank, H.\ Wang, M.\ Weides, J.\ Wenner, J.\ M.\ Martinis, and A.\ N.\ Cleland,
	Nature {\bf 464}, 697 (2010).

\bibitem{GroundStateCooling2}
	J.\ D.\ Teufel, T.\ Donner, D.\ Li, J.\ W.\ Harlow, M.\ S.\ Allman, K.\ Cicak, A.\ J.\ Sirois, J.\ D.\ Whittaker, K.\ W.\ Lehnert, and R.\ W.\ Simmonds,
	Nature {\bf 475}, 359 (2011).

\bibitem{GroundStateCooling3}
	J.\ Chan, T.\ P.\ Mayer Alegre, A.\ H.\ Safavi-Naeini, J.\ T.\ Hill, A.\ Krause, S.\ Gr{\"o}blacher, M.\ Asperlmeyer, and O.\ Painter,
	Nature {\bf 478}, 89 (2011).

\bibitem{GroundStateCooling4}
	A.\ H.\ Safavi-Naeini, J.\ Chan, J.\ T.\ Hill, T.\ P.\ Mayer Alegre, A.\ Krause, and O.\ Painter,
	Phys.\ Rev.\ Lett.\ {\bf 108}, 033602 (2012).

\bibitem{Eisert:2004aa}
	J.\ Eisert, M.\ B.\ Plenio, S.\ Bose, and J.\ Hartley,
	Phys.\ Rev.\ Lett.\ {\bf 93}, 190402 (2004).
	
\bibitem{Vitali:2007is}
	D.\ Vitali, S.\ Gigan, A.\ Ferreira, H.\ R.\ B{\"o}hm, P.\ Tombesi, A.\ Guerreiro, V.\ Vedral, A.\ Zeilinger, and M.\ Aspelmeyer,
	Phys.\ Rev.\ Lett.\ {\bf 98}, 030405 (2007).
	
\bibitem{Cavities}
	M.\ Paternostro, D.\ Vitali, S.\ Gigan, M.\ S.\ Kim, C.\ Brukner, J.\ Eisert, and M.\ Aspelmeyer,
	Phys.\ Rev.\ Lett.\ {\bf 99}, 250401 (2007).
	
\bibitem{MirrorMirrorEntanglement1}
	S.\ Mancini, V.\ Giovannetti, D.\ Vitali, and P.\ Tombesi,
	Phys.\ Rev.\ Lett.\ {\bf 88}, 120401 (2002).
	
\bibitem{MirrorMirrorEntanglement2}
	S.\ Pirandola, D.\ Vitali, P.\ Tombesi, and S.\ Lloyd,
	Phys.\ Rev.\ Lett.\ {\bf 97}, 150403 (2006).
	
\bibitem{MirrorMirrorEntanglement3}
	M.\ Pinard, A.\ Dantan, D.\ Vitali, O.\ Arcizet, T.\ Briant, and A.\ Heidmann,
	Europhys.\ Lett.\ {\bf 72}, 747 (2007).
	
\bibitem{MirrorMirrorEntanglement4}	
	M.\ J.\ Hartmann and M.\ B.\ Plenio,
	Phys.\ Rev.\ Lett.\ {\bf 101}, 200503 (2008).
	
\bibitem{Hammerer:2009tf}
	K.\ Hammerer, M.\ Aspelmeyer, E.\ S.\ Polzik, and P.\ Zoller,
	Phys.\ Rev.\ Lett.\ {\bf 102}, 020501 (2009).

\bibitem{BEC1}
	C.\ Genes, D.\ Vitali, and P.\ Tombesi,
	Phys.\ Rev.\ A {\bf 77}, 050307 (2008).
	
\bibitem{BEC2}
	G.\ De Chiara, M.\ Paternostro, and G.\ M.\ Palma,
	Phys.\ Rev.\ A {\bf 83}, 052324 (2011).

\bibitem{StefanJanJensBjoern}	
	S.\ Walter, J.\ C.\ Budich, J.\ Eisert, and B.\ Trauzettel,
	Phys.\ Rev.\ B {\bf 88}, 035441 (2013).
	
\bibitem{WalterNEMSmsb}
	S.\ Walter, T.\ L.\ Schmidt, K.\ B{\o}rkje, and B.\ Trauzettel,
	Phys.\ Rev.\ B {\bf 84}, 224510 (2011).	

\bibitem{Bolech:2007}
	C.\ Bolech and E.\ Demler,
	Phys. Rev. Lett.\ {\bf 98}, 237002 (2007).
	
\bibitem{Flowers:2007}	
	N.\ E.\ Flowers-Jacobs, D.\ R.\ Schmidt, and K.\ W.\ Lehnert,
	Phys.\ Rev.\ Lett.\ {\bf 98}, 096804 (2007)
	
\bibitem{shuttle}	
	D.\ R.\ K{\"o}nig, E.\ M.\ Weig, and J.\ P.\ Kotthaus,
	Nat.\ Nanotechnol.\ {\bf 3}, 482 (2008)
		
\bibitem{Negativity1}	
	J.\ Eisert and M.\ B.\ Plenio,
	J.\ Mod.\ Opt.\ {\bf 46}, 145 (1999).
		
\bibitem{Negativity2}
	 G.\ Vidal and R.\ F.\ Werner,
	 Phys.\ Rev.\ A {\bf 65}, 032314 (2002).
	
\bibitem{Negativity3}
         M.\ B.\ Plenio,
         Phys.\ Rev.\ Lett.\ {\bf 95}, 090503 (2005).

\bibitem{Breuer:2002wp}
	H.\ P.\ Breuer and F.\ Petruccione,
	{\it The theory of open quantum systems} (Oxford University Press, 2002).
	
	
\bibitem{Li2008}	
	T.\ F.\ Li, Yu.\ A.\ Pashkin, O.\ Astafiev, Y.\ Nakamura, J.\ S.\ Tsai. and H.\ Im,
	Appl.\ Phys.\ Lett.\ {\bf 92}, 043112 (2008).
	
\bibitem{WingreenEtAl1}
	N.\ S.\ Wingreen and Y.\ Meir, Phys.\ Rev.\ B {\bf 49}, 11040 (1994).
	
\bibitem{WingreenEtAl2}
	Y.\ Zhu, J.\ Maciejko, T.\ Ji, and H.\ Guo, Phys.\ Rev.\ B {\bf 71}, 075317 (2005).
	
\bibitem{WingreenEtAl3}
	M.-T.\ Lee and W.-M.\ Zhang, J.\ Chem.\ Phys.\ {\bf 129}, 224106 (2008).

\bibitem{Chen}
	P.\ - W.\ Chen, C.\ - C.\ Jian, and H.\ -S.\ Goan, Phys.\ Rev.\ B {\bf 83}, 115439 (2011).

\end{thebibliography}


\end{document}